\documentstyle[12pt]{article}

 2
\def\BI{{\rm 1\!l}}
\def\pois#1#2{\{#1,#2\}}
\def\Tr{\hbox{{\rm Tr}}}
\def\Eq#1{{\begin{equation} #1 \end{equation}}}
\def\blackbxx{\hbox{\leaders\hrule width 0.6em height 0.6em\hskip0.6em}}
\def\boxit{{\vbox{\hrule\hbox{\vrule
\vbox{{\phantom{\blackbxx}}
 }\vrule}\hrule}}\kern0.25em}

\newcommand{\ba}{\begin{eqnarray}}
\newcommand{\ea}{\end{eqnarray}}
\def\vLone{\matrix{{}\cr v_L \cr {}^1}}
\def\vRtwo{\matrix{{}\cr v_R \cr {}^2}}
\def\vRone{\matrix{{}\cr v_R \cr {}^1}}
\def\vLtwo{\matrix{{}\cr v_L \cr {}^2}}
\def\vaone{\matrix{{}\cr v_A \cr {}^1}}
\def\vatwo{\matrix{{}\cr v_A \cr {}^2}}
\def\dminone{\matrix{{}\cr d^{(-)} \cr {}^1}}
\def\dmintwo{\matrix{{}\cr d^{(-)} \cr {}^2}}
\def\dplustwo{\matrix{{}\cr d^{(+)} \cr {}^2}}
\def\uone{\matrix{{}\cr  u \cr {}^1}}
\def\utwo{\matrix{{}\cr u \cr {}^2}}
\def\lone{\matrix{{}\cr \ell^{(-)} \cr {}^1}}
\def\ltwo{\matrix{{}\cr \ell^{(-)} \cr {}^2}}
\def\ldminone{\matrix{{}\cr d^{(-)} \cr {}^1}(\vec{r},\hat{m}) }
\def\ldmintwo{\matrix{{}\cr d^{(-)} \cr {}^2}(\vec{r},\hat{m}) }
\def\llminone{\matrix{{}\cr \ell^{(-)} \cr {}^1}(\vec{r},\hat{m}) }
\def\llmintwo{\matrix{{}\cr \ell^{(-)} \cr {}^2}(\vec{r},\hat{m}) }
\def\luone{\matrix{{}\cr u \cr {}^1}(\vec{r},\hat{m}) }
\def\lutwo{\matrix{{}\cr u \cr {}^2}(\vec{r},\hat{m}) }
\def\ldpmone{\matrix{{}\cr d^{(\pm)} \cr {}^1}(\vec{r},\hat{m}) }
\def\ldpmtwo{\matrix{{}\cr d^{(\pm)} \cr {}^2}(\vec{r},\hat{m}) }
\def\ldplustwo{\matrix{{}\cr d^{(+)} \cr {}^2}(\vec{r},\hat{m}) }
\def\lvone{\matrix{{}\cr v \cr {}^1}(\vec{r})}
\def\lvtwo{\matrix{{}\cr v \cr {}^2}(\vec{r})}
\def\Gone{\matrix{{}\cr G^\lambda \cr {}^1}(\vec{r})}
\def\Tone{\matrix{{}\cr  T \cr {}^1}}
\def\Ttwo{\matrix{{}\cr T \cr {}^2}}
\def\Toner{\matrix{{}\cr  T \cr {}^1}(\vec{r})}
\def\Ttwor{\matrix{{}\cr T \cr {}^2}(\vec{r})}
\def\Tone{\matrix{{}\cr  T \cr {}^1}}
\def\Ttwo{\matrix{{}\cr T \cr {}^2}}
\def\Dmone{\matrix{{}\cr D^{(-)} \cr {}^1} }
\def\Dmtwo{\matrix{{}\cr D^{(-)} \cr {}^2} }
\def\Dptwo{\matrix{{}\cr D^{(+)} \cr {}^2}}
\def\LDmone{\matrix{{}\cr D^{(-)} \cr {}^1}(\vec{r},\hat{m}) }
\def\LDmtwo{\matrix{{}\cr D^{(-)} \cr {}^2}(\vec{r},\hat{m}) }
\def\LDptwo{\matrix{{}\cr D^{(+)} \cr {}^2}(\vec{r},\hat{m}) }

\begin{document}

\begin{flushright}
UAHEP 962\\
DFTUZ/96/07\\
DSFNA-T-9607\\
January 1996\\
\end{flushright}

\centerline{ \LARGE $SU_q(2)$ Lattice Gauge Theory}
\vskip 2cm

\centerline{ {\sc G. Bimonte*, A. Stern** and P. Vitale***}}

\vskip 1cm
\centerline{  The Erwin Schr\"odinger International
Institute for Mathematical Physics,}
\centerline{Pasteurgasse 6/7, A-1090 Wien, Austria}

\vskip 1cm
\centerline{*  Departamento de F\'isica Te\'orica,
Universidad de Zaragoza, 50009 Zaragoza, Spain}

\medskip

\centerline{** Dept. of Physics and Astronomy, Univ. of Alabama,
Tuscaloosa, Al 35487, U.S.A.}
\medskip

\centerline{***  Dip.to. di Scienze Fisiche, Universit\`a di Napoli,
80125 Napoli, Italy,}

\vskip 2cm

\vspace*{5mm}

\normalsize
\centerline{\bf ABSTRACT}
We reformulate the Hamiltonian approach to lattice gauge theories
such that, at the classical level, the gauge group does not act
 canonically, but instead
as a Poisson-Lie group.  At the quantum level, it then gets promoted
to a quantum group gauge symmetry.  The theory depends on two
 parameters - the deformation parameter $\lambda$ and the lattice
spacing $a$.  We show that the system of Kogut and Susskind is recovered
when $\lambda \rightarrow 0$, while QCD is recovered in the
continuum limit (for any $\lambda$).
 We thus have the possibility of having a
two parameter regularization of QCD.

\newpage
\scrollmode
\baselineskip=24pt
\section{Introduction}

There has been some interest recently in developing a q-deformed
Yang-Mills theory.\cite{qggt}  One  motivation for this activity is
the possibility of breaking the standard gauge symmetry of Yang-Mills
theories without the introduction of Higgs fields.  The attempts
  discussed  previously appear to be rather involved.  They often
require the development of differential structures on quantum groups.  In
another approach, which is the one we shall report on here, one may
q-deform  gauge theories on the lattice
- and then take the continuum limit.
However, as we shall show, this leads to  a negative result.
That is, after applying the procedure
we recover ordinary gauge theory in the continuum
limit.  Thus rather than a q-deformed
Yang-Mills theory, we have the possibility of a  two parameter
regularization of QCD, the two parameters being the deformation parameter
(which we denote by $\lambda$) and the lattice spacing $a$.  This
situation may also be of interest, since
it could then happen that certain physical quantities may converge faster in the
two-parameter space than they do for the  case of a single parameter  $a$.

Our approach is to start with the Hamiltonian formulation of lattice
gauge theories due to Kogut and Susskind.\cite{ks}  The
dynamics for that system is given in terms of   rigid rotators located at
the links of the lattice.
The  Poisson structure which is taken for the rotators is the usual one,
i.e., it is written on the cotangent bundle of the relevant group.
Recently, an alternative Poisson structure (and
  Hamiltonian) for the rigid  rotator was found.\cite{mss}  (Also see
\cite{af}.)
In the new formulation, rotations are not  canonically implemented,
but rather they are  Lie-Poisson symmetries.\cite{drin}
   Since Lie-Poisson symmetries
are known to be promoted to quantum group symmetries after quantization,
\cite{TT} the rotation group is deformed to a quantum group.
 The lattice analogue of the rotation symmetry is gauge symmetry, and
thus if we utilize this alternative Poisson structure to describe
the rigid rotators on the lattice, we obtain
 a quantum group gauge symmetry upon quantization.

 In Sec. 2 we review the standard Hamiltonian formulation of
lattice gauge theories.\cite{ks}
For simplicity we shall limit our discussion
to  $SU(2)$ gauge theories in the absence of fermions.
Then the Hamiltonian dynamics can be
written on a product space consisting
of rigid rotators, modulo the space of $SU(2)$
gauge transformations.  As stated above,  to each such rotator one
associates the standard Poisson structure, which is
written on the cotangent bundle of $SU(2)$.
Physically, a rotator is attached to each link
on the lattice, while gauge transformations
correspond to points on the lattice.
The Hamiltonian for the theory is required to
 be gauge invariant and  reduce to the
QCD Hamiltonian in the limit of zero lattice spacing, while
 gauge transformations  are required to be canonical
in the classical theory and  are
generated by Gauss law constraints.

We next  review the alternative
 classical Hamiltonian formulation of the rigid
 rotator\cite{mss}, \cite{af} in  Sec. 3.   There  the six dimensional
 phase space is taken to be the $SL(2,C)$ group manifold.   As in
the standard formulation, the Hamiltonian is  invariant  under
 left and right $SU(2)$ transformations
  (which contain ordinary rotations).  Once again, these
transformations do not correspond to canonical symmetries, but rather
they  are  Lie-Poisson symmetries.  In the system given here
we introduce a parameter $\lambda$ (the `deformation parameter'),
and the standard Hamiltonian
formalism for the rotator is recovered
in the limit $\lambda \rightarrow 0$.

The alternative Hamiltonian formulation of the rigid rotator
is  applied to lattice gauge
theories in Sec. 4.   We thereby deform the Kogut Susskind Hamiltonian
dynamics in a way
which preserves the $SU(2)$ gauge symmetry of the classical theory
  - but which replaces    canonical symmetry transformations
by  Lie-Poisson transformations.  As in the Kogut Susskind system,
 rotator degrees of freedom are assigned to each link on the lattice
and  gauge transformations are associated with each point
 on the lattice.  However, now we shall describe
the rotator degrees of freedom in terms of $SL(2,C)$ variables.
 In addition to the parameter $a$ denoting the lattice spacing,
our theory possesses the deformation parameter $\lambda$  and
we require that the Kogut Susskind Hamiltonian formalism is
recovered in the limit   $\lambda\rightarrow 0$.
We shall show that our system yields the correct continuum limit
of $SU(2)$ gauge theory even when the deformation parameter is
different from zero.

In section  5 we make some preliminary remarks about the quantization
of this system.       When Lie-Poisson symmetries are
present in the classical theory
the standard practice is to apply the method
 of deformation quantization  \cite{bffls}.  Fixing  the quantum
dynamics using the method of deformation quantization requires
writing down a star product on the space of classical observables.
  This is in general a difficult task and shall not
be attempted here.  Instead, we shall only demand, as is usually done,
 that the $SU(2)$ Poisson Lie group  symmetry of the
classical theory gets replaced upon quantization
by an $SU_q(2)$ quantum group  symmetry\cite{TT}.
Since an $SU_q(2)$ matrix is attached
at each point  on the lattice,  we in fact
end up with an $SU_q(2)$ gauge  symmetry.  The quantum analogues of the
$SL(2,C)$ matrices are  quantum double matrices which are then associated
with each link on the lattice.
The commutation relations for  the latter matrices are required to be
 covariant under $SU_q(2)$ gauge transformations, while the quantum
Hamiltonian is invariant.

After this work was completed we learned of a series of papers
 by S.A. Frolov\cite{fro} where a similar system is discussed.  It
differs from ours in the nature of gauge transformations.  For us,
all link variables transform in an identical fashion at any particular
site, as in the spirit of Kogut and Susskind.  As a result of this it
is easy to write down the Wilson loop operators, but difficult to give an
explicit expression for the gauge generator (i.e., the Gauss law) in
terms of the link variables.  (So far we have been unable to do it.)
On the other hand, Frolov simply postulates a Gauss law.  From it
one then obtains a complicated gauge transformation rule for the
link variables (where links attached to same vertex transform
differently), and hence a complicated form for the Wilson loop operators.

\section{Review of the Kogut Susskind Hamiltonian Formalism}

As stated earlier, we limit our discussion
to  $SU(2)$ gauge theories in the absence of fermions, and the
  Hamiltonian formulation of
lattice gauge theories\cite{ks} is then described solely in terms of
rigid rotators.  In the standard formulation
the phase space for a single rotator
 is spanned by angular momentum variables $j_a$ and an $SU(2)$
matrix $u$ with Poisson brackets given by
 \ba
{\pois{j_a}{j_b}}& =&\epsilon_{abc}\;j_c   \;,
\label{canb1} \\
{\pois{j_a}{u}}& =&  \frac i2 \sigma_a u  \;,
\label{canb2}
\ea
with the Poisson brackets of matrix elements of
$u$ with themselves being zero.   $\sigma_a$ are Pauli matrices.
The brackets for $u$ and $j_a$
are preserved under the following transformation:
\ba  u&\rightarrow &v_L^\dagger\;  u\; v_R \;,\label{cant1}\\
j_a\sigma_a& \rightarrow &v_L^\dagger\; j_a\sigma_a \;
 v_L\;,\label{cant2} \ea
where $v_L$ and $ v_R$ are independent $SU(2)$ matrices.  It easily
follows that  (\ref{cant1}) and (\ref{cant2}) are canonical
transformations.

Physically, we are to suppose that a rotator [with Poisson brackets
given by (\ref{canb1}) and (\ref{canb2})] is attached to each link
on a three dimensional cubic lattice, while gauge transformations
[analogous to those in
 (\ref{cant1}) and (\ref{cant2}) parametrized by $v_L$ and $ v_R$]
are associated with the points on the lattice.
  We  make this statement more precise below.

 Following   ref. \cite{ks},
an arbitrary point on a lattice will be denoted by
a vector $\vec{r}$, while a link connecting point
 $\vec{r}$ to a neighboring point in the direction
  $ \hat{m}$ is denoted by  ($\vec{r}, \hat{m})$.  Here $\hat{m}$
can be one of the six unit vectors running along the lattice.
For each link $(\vec{r}, \hat{m})$ we have the phase space variables
\Eq{u(\vec{r}, \hat{m})\in SU(2)\quad{\rm and}\quad
j_a(\vec{r}, \hat{m})\;.\label{ujvar}}
The inverse of
$u(\vec{r}, \hat{m})$ is obtained by traversing the link
$(\vec{r}, \hat{m})$ in the $-\hat{m}$ direction, i.e.
\Eq{u(\vec{r}, \hat{m})^\dagger =  u(\vec{r}+a\hat{m}, - \hat{m})\;.
\label{invu}}
The phase space for  $SU(2)$ lattice gauge theory
 consists of all variables (\ref{ujvar}).
Their nonvanishing Poisson brackets are given by
 \ba    \pois{j_a(\vec{r}, \hat{m})}  {j_b(\vec{r}, \hat{m})}
& =&\epsilon_{abc}\;j_c(\vec{r}, \hat{m})
\label{canbrabp} \\
\pois{j_a(\vec{r}, \hat{m}) }{u(\vec{r}, \hat{m})}
& =&  \frac i2 \sigma_a\; u(\vec{r}, \hat{m})
  \;,          \label{canbraep}    \ea
for all links $(\vec{r}, \hat{m})$.

The Poisson brackets (\ref{canbrabp}) and (\ref{canbraep}) must
yield the standard Poisson structure for Yang-Mills theory in the
continuum limit, i.e.
\Eq{\pois{A_i^a(\vec{x})}{E_j^b(\vec{y})}
= \delta_{ij}\delta_{ab}\delta^3(\vec{x}-\vec{y})\;,
\label{YMps}} where $A_i^a(\vec{x})$ are Yang-Mills potentials and
$E_j^b(\vec{y})$ are the electric field strengths.
In order to recover this bracket from
 (\ref{canbrabp}) and (\ref{canbraep}),  one interprets $
 u(\vec{r}, \hat{m})$ according to
 \Eq{u(\vec{r}, \hat{m}) =\exp( iag
\frac{\sigma_a}{2} A_i^a(\vec{r})\hat{m}_i  )\;.\label{will}}
Then $u(\vec{r},\hat{m})$ goes to $1+iag
\frac{\sigma_a}{2} A_i^a(\vec{r})\hat{m}_i$
when $a\rightarrow 0$.  $j_a(\vec{r}, \hat{m})$ is interpreted
as the line integral of the electric field along the link
 $(\vec{r}, \hat{m})$.  Thus
\Eq{j_a(\vec{r}, \hat{m})=-\frac{a^2}g E_i^a(\vec{r})\hat{m}_i\;.
\label{jiE}}
Using (\ref{canbraep}) we then get
\Eq{ \pois{A_i^a(\vec{r})} { E_j^b(\vec{r}) } \rightarrow
\frac1{a^3} \delta_{ij}\delta_{ab}\;,\quad {\rm as}\;
  a\rightarrow 0    \;, \label{pbAE}}
  which agrees with (\ref{YMps}).  [From
 (\ref{canbrabp}) the Poisson brackets of
$ E(\vec{r})$ with itself go like $1/a^2$ which as a density
distribution vanishes in the continuum limit.]

Gauge transformations  are associated with points on the lattice.
At the point  $\vec{r}$ we define
$$v(\vec{r})\in SU(2)\;.$$    As we are in the Hamiltonian formulation
of the theory the gauge transformations are time independent.
Gauge transformations on the phase
space variables $u(\vec{r}, \hat{m})$  and $j_a(\vec{r}, \hat{m})$
correspond to
\ba  u(\vec{r}, \hat{m}) &\rightarrow &v(\vec{r})^\dagger
\;  u(\vec{r}, \hat{m}) \; v(\vec{r}+a\hat{m}) \;,
 \label{lgt}     \\
j_a(\vec{r}, \hat{m})\sigma_a& \rightarrow &v(\vec{r})^\dagger\; j_a
(\vec{r}, \hat{m})\sigma_a \;  v(\vec{r}) \; \;.
 \label{lgt2}  \ea
The transformation (\ref{lgt}) is consistent with
(\ref{invu}).
Upon comparing with (\ref{cant1}) and (\ref{cant2}) it is evident
that  (\ref{lgt}) and  (\ref{lgt2}) are canonical transformations.

The next task is to write down the Hamiltonian.  The requirements
are that it be gauge invariant and also that it reduces to the
standard field theory
Hamiltonian in the limit of zero lattice spacing.   Concerning
the first requirement, an obvious
set of  gauge invariant quantities are the kinetic
energies of the  rotators.
Indeed, the sum of all such kinetic energies is
one ingredient $H_0$ in the Kogut Susskind Hamiltonian,
\Eq{ H_0 = \frac{g^2}{2a} \sum_{\vec{r},\hat{m}>0}
j_a(\vec{r}, \hat{m})   j_a(\vec{r}, \hat{m})   \;,\label{hzero}}
where $\hat{m}>0$ indicates that the sum is only over positive
directions of  $\hat{m}>0$.
The constants $g$ and $a$ represent the gauge coupling and lattice
spacing, respectively.  The Hamiltonian (\ref{hzero})
cannot be the full
story since it leads to a trivial system with all rotators
noninteracting.  Furthermore,
in the limit of zero lattice spacing, $H_0$
only gives the electric field contribution to the QCD Hamiltonian.
It is well known that the magnetic field contribution can
come from Wilson loop variables constructed
on the lattice.  For this we
let $\Gamma_{(\vec{r}, \hat{m},\hat{n})} $
denote a square plaquette connecting the point $\vec{r}$ to its nearest
neighbors in the lattice $\hat{m}$ and $\hat{n} $ directions and then
denote the associated Wilson loop by  $W(
\Gamma_{(\vec{r}, \hat{m},\hat{n})})$.  It is given by
\Eq{ W( \Gamma_{(\vec{r}, \hat{m},\hat{n})})=
\Tr\;u(\vec{r}, \hat{m})\; u(\vec{r}+a\hat{m}, \hat{n}) \;
u(\vec{r}+a \hat{n},\hat{m})^\dagger u(\vec{r},\hat{n})^\dagger
\;. \label{wl}}
From (\ref{lgt}) it is clear that (\ref{wl}) is gauge invariant for all
 $\vec{r}$, $\hat{m}$ and $\hat{n} $.  It is also clear that terms
like (\ref{wl}) introduce nontrivial interactions between the
rotators.  Upon writing $u(\vec{r}, \hat{m})$ according to
(\ref{will}),   it has been shown
that  $W(\Gamma_{(\vec{r}, \hat{m},\hat{n})})$ yields the
usual magnetic field
contribution to the action [associated with the
plaquette  $\Gamma_{(\vec{r}, \hat{m},\hat{n})}$]
 upon taking the continuum  limit $a\rightarrow 0$.
More specifically, upon expanding to fourth order in $a$ we get
that  \Eq{
W(\Gamma_{(\vec{r}, \hat{m},\hat{n})})\rightarrow \Tr \biggl(
\BI  - \frac12  g^2 a^4 (F_{ \hat{m}\hat{n}}(\vec{r}))^2 \biggr)
 \;,\quad {\rm as}  \;   a\rightarrow 0\;,\label{tsr}} where
$ F_{ \hat{m}\hat{n}}(\vec{r})=  F_{ij}^a(\vec{r})\frac{\sigma_a}2
 \hat{m}_i\hat{n}_j$, $F_{ij}^a=\epsilon_{ijk} B^a_k$
being the magnetic field strength tensor.
We can finally express the Kogut Susskind Hamiltonian according to
\Eq{H=H_0+H_1\;,\qquad H_1=\frac 1{ag^2} \sum_{\boxit}
\biggl( W( \Gamma_{(\vec{r}, \hat{m},\hat{n})})+
 W( \Gamma_{(\vec{r}, \hat{m},\hat{n})})^* -4\biggr)
 \;.\label{ksh}}
The sum in $H_1$ is over all plaquettes.
The coefficients in $H_0$ and $H_1$ were chosen so that $H$ has
the usual continuum limit, i.e.
  \Eq{     H\rightarrow   \frac12  \sum_{\vec{r}} \;a^3 \biggl(
E^a_i(\vec{r}) E^a_i(\vec{r}) +
B^a_i(\vec{r}) B^a_i(\vec{r}) \biggr)\;,\quad{\rm
as}\;a\rightarrow 0\;,}

A final ingredient in this system (which turns out to be a source
of difficulty for us when we deform the system)
is that the gauge symmetry
(\ref{lgt}) and (\ref{lgt2}) implies the existence of first
class constraints.  These constraints
$G_a(\vec{r}) \approx 0$  are defined at points on the lattice
and they generate the gauge symmetry.  Thus their nonvanishing
 Poisson brackets may be given by:
 \ba    \pois{G_a(\vec{r})}  {j_b(\vec{r}, \hat{m})}
& =&\epsilon_{abc}\;j_c(\vec{r}, \hat{m})   \\
 \pois{G_a(\vec{r})}{u(\vec{r}, \hat{m})}
&=& \frac i2 \sigma_a\; u(\vec{r}, \hat{m}) \;. \ea
A solution for $G_a(\vec{r}) $  is \Eq{G_a(\vec{r}) = \sum_{\hat{m}}
j_a(\vec{r}, \hat{m})  \;,}the sum being over all links connected to
the lattice point $\vec{r}$.
In the continuum limit this gives the usual Gauss law constraint.
In the quantum theory we must impose that the operator
corresponding to $G_a(\vec{r}) $ annihilates
all physical states.

\section{Alternative Hamiltonian formulation of the Rigid Rotator}

We next review the alternative
 classical Hamiltonian formulation of the rotator given in
ref. \cite{mss}.  There it was shown that the six dimensional
phase space describing a rigid rotator could be spanned
by the set $\{d^{(-)}\}$ of $2\times 2$
complex unimodular matrices which constitute
 `classical double variables'.  Thus the
 phase space is  $SL(2,C)$.

The Hamiltonian of ref. \cite{mss} was a function of only
  $\Tr d^{(-)} { d^{(-)}}^\dagger$.  More specifically, it was given
by  \Eq{     H_{rot}(\lambda)={1\over {2\lambda^2}}
(\Tr \; d^{(-)}{d^{(-)}}^\dagger\;-\;2) \;,\label{ham} }
where $\lambda $ is a constant which plays the role of a deformation
parameter.   This Hamiltonian is  invariant  under
 left and right $SU(2)$ transformations:
\Eq{ d^{(-)} \rightarrow v_L^\dagger\; d^{(-)}\; v_R\;,\quad v_L
,v_R\in SU(2)\;,   \label{gtr}}          and
these transformations are analogous to separate rotations of the
body and space axes of a rigid rotator.

  Although the Hamiltonian is invariant under (\ref{gtr}), these
transformations do not correspond to canonical symmetries due
to a nontrivial quadratic Poisson structure which was assumed for
$d^{(-)}$.  Using tensor product notation the Poisson brackets can
be compactly written as follows:
\Eq{ \{\dminone,\dmintwo\}=-\dminone \dmintwo\;r\;-\;
r^\dagger\;\dminone
\dmintwo\; \;. \label{pbdp} }
Here $\dminone,\dmintwo$ and $r$ denote $4 \times 4$ matrices with
$\dminone =d^{(-)}\otimes \BI$, $\dmintwo =\BI \otimes d^{(-)}$ and
$r$ given by
\Eq{ r={{i\lambda}\over 4}\pmatrix{1 & & &  \cr &-1 & &  \cr &4 &-1& \cr
& & & 1\cr} \quad \label{rmat}\;.}
  The brackets (\ref{pbdp}) are skew symmetric.  Furthermore, the
  $r-$matrix (\ref{rmat}) is known to satisfy the classical
Yang-Baxter relations which insures that the Jacobi identity holds for
(\ref{pbdp}).  It is clear that the transformations
(\ref{gtr}) do not preserve the Poisson brackets and hence they are
not canonical symmetries.
They are instead Lie-Poisson symmetries.  For this
 we now associate a certain Poisson structure
to the $SU(2)$ matrices $v_L$ and $v_R$ involved in the transformations
 (\ref{gtr}).
These Poisson brackets are chosen to be compatible with those of
the observables $d^{(-)}$ [cf. eq.  (\ref{pbdp}) ]  along with
the transformation (\ref{gtr}).  They are  given by
\ba \{\vaone,\vatwo\}&=&[\;r\;,\;\vaone \vatwo\;]  \;,\qquad A=L,R
\label{pbva} \\
\{\vLone,\vRtwo\}&=&0  \;,   \\
\{\vaone,\dmintwo\}&=&0  \;,   \label{pbvd}      \ea
where $\vaone =v_A\otimes \BI$, and $\vatwo =\BI \otimes v_A$.
To show compatibility we note that the left hand side of
(\ref{pbdp}) transforms to
\Eq{\pois {\vLone^\dagger \dminone \vRone}
 {\vLtwo^\dagger \dmintwo \vRtwo} \;,} which we now can
 compute using (\ref{pbdp}) and (\ref{pbva}-\ref{pbvd}).  We get
\Eq{ -\vLone^\dagger
\dminone\vRone \vLtwo^\dagger\dmintwo\vRtwo\;r\;
-\;r^\dagger\;\vLone^\dagger\dminone\vRone
\vLtwo^\dagger\dmintwo\vRtwo\; \;. \label{trpbdp} }  Using
(\ref{gtr}), the right hand side of (\ref{pbdp}) also transforms to
(\ref{trpbdp}), thus showing that the brackets
 (\ref{pbva}-\ref{pbvd}) for $v_A$ are compatible with
 (\ref{pbdp}) and the transformation (\ref{gtr}).
Consequently (\ref{gtr}) is a Lie-Poisson  transformation.

We next proceed with the Hamilton equations of motion.  For this,
the Poisson brackets  (\ref{pbdp})  are insufficient because
the Hamiltonian for the system involves ${d^{(-)}}^\dagger$, as well
as $d^{(-)}$.  We therefore need to know
the Poisson brackets of ${d^{(-)}}^\dagger$ with
$d^{(-)}$.  From \cite{mss} we have  \Eq{ \{ \dminone,\dplustwo \}=
-\dminone \dplustwo \; r-r \;\dminone\dplustwo\;,
\label{pbdpm}} where $d^{(+)}={{d^{(-)}}^\dagger}^{-1}$.
The variables $d^{(\pm)}$ along with the Poisson brackets
(\ref{pbdp}) and (\ref{pbdpm})
(which are consistent with the Jacobi identity)
define the {\it classical double}.
Using eqs. (\ref{ham}), (\ref{pbdp}) and (\ref{pbdpm}) one obtains
the following equations of motion.  They were found to be:
\Eq {\frac d {dt} d^{(-)}\; {d^{(-)}}^{-1} =\frac i{2\lambda}
[ d^{(-)} {d^{(-)}}^\dagger ]_{t\ell }\;,\label{eom}}
where $[A]_{t\ell}$ denotes the traceless part of $2\times 2$ matrix
 $A$, i.e.,  $[A]_{t\ell}=A-\frac 12 \Tr (A) \times \BI$.

To make the connection with the isotropic rigid rotator system,
 we  decompose the $SL(2,C)$ matrix $d^{(-)}$
into the product of an element of the
$SU(2)$ subgroup and an element of the Borel subgroup $SB(2,C)$.
We do this as follows:
\Eq{ d^{(-)}=\ell^{(-)} u\;,\qquad u\in SU(2)\;, \quad
\ell^{(-)} =\pmatrix{x_0 & 0  \cr x_- & x_0^{-1} \cr} \in SB(2,C)\;,
\label{deg}}
 where $x_0$ is real and $x_-$ is complex.  The map from
$ SU(2)\times    SB(2,C)$ to $SL(2,C)$ is two-to-one.
 The Poisson brackets
(\ref{pbdp}) for the classical double variables $d^{(-)}$ are recovered
with the following choice of brackets for $u$ and $\ell^{(-)}$:
\ba \{\uone,\utwo\}&=&[\;r\;,\;\uone \utwo\;]  \;, \label{pblub}\\
\{\lone,\ltwo\}&=&-[\;r\;,\;\lone \ltwo\;]  \\
\{\lone,\utwo\}&=&-\lone\; r \;\utwo\;.\label{pblue} \ea
Upon substituting (\ref{deg}) into the equation of motion
 (\ref{eom}) we get
 \Eq {\frac d {dt} u\; u^\dagger -\frac i{2\lambda}
[{ \ell^{(-)}}^\dagger \ell^{(-)}]_{t\ell } = -
{\ell^{(-)}}^{-1} \; \frac d {dt} \ell^{(-)}
\;.\label{eom2}}
The left hand side of eq. (\ref{eom2}) is traceless and antihermitean,
while the right hand side is an element of the
$\underline{SB(2,C)}$ Lie algebra.  It follows that the left and
right hand sides must separately vanish, leading to:
 \Eq{\frac d {dt} u\; u^\dagger   = \frac i2 J_a \sigma_a\;, \qquad
{\rm where} \qquad   J_a = \frac 1{2\lambda} \Tr \;
{ \ell^{(-)}}^\dagger \ell^{(-)}\sigma_a \;,\label{eom3}}
along with    \Eq{\frac d {dt} \ell^{(-)} =0 \;.\label{eom4}}
$J_a$ in eq. (\ref{eom3})
can now be interpreted as the physical
angular momenta of the isotropic
rotator.  Since they are functions of only $\ell^{(-)}$ and
 ${\ell^{(-)}}^\dagger$,  they are conserved.  Now if
we associate the $SU(2)$ matrix
$u$ with the orientation of the rotator (or the transformation
between space and body axes), then eq. (\ref{eom3}) gives the
desired result, namely,
 that the rigid body undergoes a uniform precession.

The above Hamiltonian formulation of the isotropic rotator is a
deformation of the usual Hamiltonian formulation,
where as we stated earlier,
 $\lambda $ plays the role of the deformation
parameter.   The usual Hamiltonian formulation is recovered when
$\lambda$ goes to zero, which we refer to as the `canonical limit'.
 For this to happen we must first
show how to express the matrix $\ell^{(-)}$ in terms of the
 canonical angular momentum variables $j_a$ [as opposed to the
variables  $J_a$
which do not obey the canonical Poisson bracket relations
(\ref{canb1})].    We write
\Eq{ \ell^{(-)} = \exp{(i\lambda e^a j_a})\label{lioj}\;,}
 where $e^a$ are generators of $SB(2,C)$.  They
can be expressed in terms of Pauli matrices $\sigma_a$ as follows:
\Eq{  e^a ={1\over 2} (i\sigma_a +\epsilon_{ab3} \sigma_b)\;. }
(Actually, for the purpose of taking the canonical
limit, we only need that  relation (\ref{lioj})
holds up to second order in
$\lambda$.)  Eq. (\ref{lioj}) is equivalent to assigning
$x_0$ and $x_-$ in (\ref{deg}) according to
\Eq{x_0=\exp{\biggl(-\frac{\lambda j_3}2\biggr)} \;,\qquad
x_-=-2\;\frac{j_1+ij_2}{j_3}\sinh{\biggl(\frac
{\lambda j_3}2\biggr)} \;.\label{defxs}}
Now to lowest order in $\lambda$, the Hamiltonian (\ref{ham})
reduces to \Eq{H_{rot}(\lambda\rightarrow 0)
=\frac12\; j_aj_a\;,\label{ch}}
while, using eq. (\ref{deg}), the Poisson brackets (\ref{pblub}-
\ref{pblue}) go to
 \ba
{\pois{j_a}{j_b}}_{\lambda\rightarrow 0}& =&\epsilon_{abc}\;j_c   \;,
\label{canbrab} \\
{\pois{j_a}{u}}_{\lambda\rightarrow 0}& =&  \frac i2 \sigma_a u    \;,\\
\{\uone,\utwo{\}}_{\lambda\rightarrow 0}& =& 0   \;.
\label{canbrae}
\ea
which agrees with eqs. (\ref{canb1}) and (\ref{canb2}).
We thus arrive at the canonical description of an isotropic rotator
with the moment of inertia set equal to one.  The Hamiltonian
 (\ref{ham}) and the Poisson brackets (\ref{pblub}-\ref{pblue})
therefore define a one parameter deformation of the canonical
description of the  isotropic rigid rotator.  When $\lambda \ne
0$, the chiral transformations are Lie-Poisson symmetries.
They  reduce to canonical
symmetries when $\lambda \rightarrow 0$.  In that limit, the variables
$u$ and $j_a$ transform in the usual way, i.e. as in
(\ref{cant1}) and (\ref{cant2}).

\section{Deformation of $SU(2)$ lattice gauge theory}

We are now ready to deform the Kogut Susskind Hamiltonian
dynamics in a way
which preserves the $SU(2)$ gauge symmetry of the classical
 Hamiltonian function - but which replaces
 the  canonical symmetry transformations
into  Lie-Poisson transformations.  As before we assign rotator
degrees of freedom to each link on the lattice.
 However, now we shall describe
the rotator degrees of freedom in terms of  classical double variables
\Eq{ d^{(-)}(\vec{r}, \hat{m})\in SL(2,C)\;,\label{dvar}}
rather than the variables
$ j_a(\vec{r}, \hat{m})$ and $u(\vec{r}, \hat{m})$.

  To each classical double variable $ d^{(-)}(\vec{r}, \hat{m})$
 we assign the Poisson structure given by (\ref{pbdp}) and (\ref{pbdpm}).
Thus we replace the set of
 Poisson brackets (\ref{canbrabp}) and (\ref{canbraep}) by
\ba\{\ldminone,\ldmintwo\}&
=&-\ldminone \ldmintwo\;r-r^\dagger\;\ldminone
\ldmintwo\;   , \label{pbldp} \\
 \{ \ldminone,\ldplustwo \}&=&
-\ldminone \ldplustwo \; r-r \;\ldminone\ldplustwo\;,
\label{pbldpm}\ea for all links $(\vec{r}, \hat{m})$, and
 where \Eq{d^{(+)}(\vec{r}, \hat{m})
={{d^{(-)}}(\vec{r}, \hat{m})^\dagger}^{-1}\;.\label{dpid1}  }
  [In analogy
to (\ref{invu}), we obtain the inverse of
$d^{(-)}(\vec{r}, \hat{m})$ by traversing the link
$(\vec{r}, \hat{m})$ in the $-\hat{m}$ direction, i.e. \Eq{{d^{(-)}
(\vec{r}, \hat{m})}^{-1} =d^{(-)}(\vec{r}+a\hat{m}, - \hat{m})\;\;.
\quad ]\label{dpid2}}
In (\ref{pbldp}) and (\ref{pbldpm}), we once again
resort to tensor product notation with
$\ldpmone=d^{(\pm)}(\vec{r}, \hat{m})\otimes \BI$,
$\ldpmtwo=\BI \otimes d^{(\pm)}(\vec{r}, \hat{m})$  and the $r$-matrix
 is given by (\ref{rmat}).  Since the $r-$matrix depends on
$\lambda$ we are introducing a new parameter (in addition to $a$)
to the lattice theory.  We
shall require that the Kogut Susskind Hamiltonian formalism is
recovered in the limit
 $\lambda\rightarrow 0$.  This is in addition to the
requirement that the Hamiltonian function be gauge invariant.

With regard to gauge invariance, for us
gauge transformations are associated with points on the lattice,
just as was the case in the Kogut Susskind formalism.  Thus
at the point  $\vec{r}$ we once again define
$v(\vec{r})\in SU(2)\;.$   Now in analogy to
 (\ref{gtr}), we uniquely define
 gauge transformations of the phase space variables
$d^{(\pm)}(\vec{r}, \hat{m})$  according to
\Eq{ d^{(\pm)}(\vec{r}, \hat{m})   \rightarrow v(\vec{r})^\dagger  \;
 d^{(\pm)}(\vec{r}, \hat{m})
 \; v(\vec{r}+a\hat{m}) \;.   \label{lgtr}}
As was true for a single rotator, such transformations are not
canonical.  Instead they are Lie Poisson.  In this regard, we attach
a product Poisson structure to the gauge degrees of freedom
$v(\vec{r}),$
\Eq{ \{\lvone,\lvtwo\} = [\;r\;,\;\lvone \lvtwo\;]  \;,\label{pbvrvr}}
This Poisson structure is compatible with the brackets
(\ref{pbldp}) and (\ref{pbldpm}), along with the transformations
(\ref{lgtr}).  Hence (\ref{lgtr}) are Lie-Poisson transformations.

From (\ref{ham}) we already know how to write down the gauge invariant
deformation of  the first term $H_0$ in the
Kogut Susskind Hamiltonian.  It is just
 \Eq{     H_{0}(\lambda)={{g^2}\over {2a\lambda^2}}
 \sum_{\vec{r},\hat{m}>0}
\biggl( \Tr \; d^{(-)}(\vec{r}, \hat{m})
 \; d^{(-)}(\vec{r}, \hat{m})^\dagger\;-\;2\biggr)    \;.\label{lham} }
We can make a decomposition of the $SL(2,C)$ matrix
$ d^{(-)}(\vec{r}, \hat{m}) $ in terms of $SU(2)$ and $SB(2,C)$
subgroups in an identical manner to what was done in the previous
section, i.e.
\Eq{ d^{(-)}(\vec{r}, \hat{m})
=\ell^{(-)}(\vec{r}, \hat{m})\;u(\vec{r}, \hat{m})
 \;,\qquad u(\vec{r}, \hat{m}) \in SU(2)\;, \quad
\ell^{(-)}(\vec{r}, \hat{m}) \in SB(2,C) \;.\label{decd}}
 The Poisson brackets
(\ref{pbldp}) for the classical double variables $d^{(-)}
(\vec{r}, \hat{m})$ are recovered
with the following choice of brackets for $u(\vec{r}, \hat{m})
$ and $\ell^{(-)}(\vec{r}, \hat{m})$:
\ba \{\luone,\lutwo\}&=&[\;r\;,\;\luone \lutwo\;]  \;, \label{pbllub}\\
\{\llminone,\llmintwo\}&=&-[\;r\;,\;\llminone \llmintwo\;]  \\
\{\llminone,\lutwo\}&=&-\llminone\; r \;\lutwo\;.\label{pbllue} \ea
  Furthermore, we make the $SB(2,C)$ matrix
 depend on $\lambda$ as in the previous section, i.e.
\Eq{ \ell^{(-)}(\vec{r}, \hat{m})
= \exp{\biggl(i\lambda e^a j_a(\vec{r}, \hat{m})}\biggr)\;.
\label{lioj2}}
  It then follows that $     H_{0}(\lambda)$ is a deformation of
 $H_0$, i.e. in analogy to (\ref{ch}),  we have
$$     H_{0}(\lambda\rightarrow 0)=H_0\;.$$
Also the transformation (\ref{lgtr}) reduce to
 canonical transformations (\ref{lgt}) and  (\ref{lgt2}) in the limit.

It is also straightforward to  write down the gauge invariant
deformation of the second term
$H_1$ in the Kogut Susskind Hamiltonian.
 For this we construct a set of Wilson loop
variables   $W^\lambda(\Gamma_{(\vec{r}, \hat{m},\hat{n})})$ using
$ d^{(-)}(\vec{r}, \hat{m})$,
\Eq{ W^\lambda( \Gamma_{(\vec{r}, \hat{m},\hat{n})})=
\Tr \;d^{(-)}(\vec{r}, \hat{m})\;
 d^{(-)}(\vec{r}+a\hat{m}, \hat{n}) \;
d^{(+)}(\vec{r}+ a\hat{n},\hat{m})^\dagger \;
 d^{(+)}(\vec{r},\hat{n})^\dagger   \;. \label{dwl}}
From (\ref{lgtr}) it easily follows that
$W^\lambda(\Gamma_{(\vec{r}, \hat{m},\hat{n})})$ is gauge invariant.
In this regard (\ref{dwl}) is not unique, as we can arbitrarily
replace the different
factors $d^{(\pm)}$ with $d^{(\mp)}$ in the formula for
$ W^\lambda( \Gamma_{(\vec{r}, \hat{m},\hat{n})})$.
On the other hand we shall show that for the choice (\ref{dwl})
we recover the correct continuum limit ($a\rightarrow 0 $)
of $SU(2)$ gauge theory.  Concerning the canonical
limit $\lambda\rightarrow 0$,
upon using (\ref{decd}) and (\ref{lioj2}), we get that
$ d^{(\pm)}(\vec{r}, \hat{m}) \rightarrow
 u(\vec{r}, \hat{m})$  and hence
$$W^{\lambda\rightarrow 0}(\Gamma_{(\vec{r}, \hat{m},\hat{n})})=
 W(\Gamma_{(\vec{r}, \hat{m},\hat{n})})\;.$$
  Thus $W^\lambda(\Gamma_{(\vec{r}, \hat{m},\hat{n})})$
   is a gauge invariant
 deformation  of $W(\Gamma_{(\vec{r}, \hat{m},\hat{n})})$.
We can now write down the gauge invariant deformation of the Kogut
Susskind Hamiltonian.  It is
\Eq{H(\lambda)=H_0(\lambda) +H_1(\lambda)
\;,\qquad H_1(\lambda)
=\frac 1{ag^2}\sum_{\boxit}
\biggl( W^\lambda( \Gamma_{(\vec{r}, \hat{m},\hat{n})})+
 W^\lambda( \Gamma_{(\vec{r}, \hat{m},\hat{n})})^* -4\biggr)
 \;.\label{dksh}}
The sum in $H_1$ is over all plaquettes.

The final ingredient in this system
is the analogue of the Gauss law constraints
$G^\lambda_a(\vec{r}) \approx 0$.  They generate the gauge symmetry
(\ref{lgtr})  and thus their non
vanishing Poisson brackets should be of the form:
 \Eq{ \pois{\Gone}{\ldpmtwo}=  X(\vec{r}) \ldpmtwo\;. }
We want that $G^\lambda(\vec{r})\rightarrow G_a(\vec{r})
\frac{\sigma_a}2$ when
$\lambda \rightarrow 0$.  We have been unable to find an
explicit solution for $G^\lambda(\vec{r})$
expressed in terms of
$\ell^{(-)}(\vec{r}, \hat{m})$ and $u(\vec{r}, \hat{m}) $
which is consistent with these requirements.

Above we have shown that the canonical
limit ($\lambda\rightarrow 0$) of our model is the system of Kogut
Susskind.   We now show that the continuum limit
 ($a\rightarrow 0 $) of our model is standard
 $SU(2)$ gauge theory (for
any value of $\lambda$).  In this regard, we once again write
$u(\vec{r}, \hat{m}) $ in terms of Yang-Mills potentials
$A_i^a(\vec{x})$  according to
(\ref{will}).  In addition, using (\ref{jiE}) and
(\ref{lioj}) we get that  \Eq{ \ell^{(-)} (\vec{r}, \hat{m})
= \exp{\biggl(-i\lambda \frac{a^2}ge^a E_i^a(\vec{r})\hat{m}_i \biggr)
}\;.   \label{lliE}}  Now from the Poisson bracket
(\ref{pbllue}) we get that
\Eq{\pois{E^a_i(\vec{r})}{A^b_j(\vec{r})} \;e^a\otimes\sigma^b
\rightarrow - \frac2{\lambda a^3} \delta_{ij}  \;r  \;,}
when $a \rightarrow 0 $.  Because
 the classical $r-$matrix can be written
$r=\frac{\lambda}2 e^a\otimes \sigma^a$, we recover the canonical
Poisson brackets (\ref{pbAE}) for $SU(2)$ gauge theory.
[These are the only nonvanishing Poisson brackets, as
$\pois{E}{E}$ and $\pois{A}{A}$ go like
$a^n, \; n>-3$, which as a density
distribution vanishes in the continuum limit.]

It remains to take the continuum  limit of the
Hamiltonain $H(\lambda)$, eq. (\ref{lham}).
 In this regard we note that the term  $H_0(\lambda)$ can be written
 \Eq{     H_{0}(\lambda)={{g^2}\over {2a\lambda^2}}
 \sum_{\vec{r},\hat{m}>0}
\biggl( \Tr \; \ell^{(-)}(\vec{r}, \hat{m})
 \;\ell^{(-)}(\vec{r}, \hat{m})^\dagger\;-\;2\biggr)    \;. }
Then using (\ref{lliE}) we find that
$ H_{0}(\lambda)$ yields the electric field energy of $SU(2)$
gauge theory   \Eq{     H_{0}(\lambda)\rightarrow   \frac12
\sum_{\vec{r}} \;a^3E^a_i(\vec{r}) E^a_i(\vec{r}) \;,}
as $a\rightarrow 0$.  Furthermore,
$ H_{1}(\lambda)$  in eq. (\ref{dksh})
yields the magnetic field energy of $SU(2)$
gauge theory but this requires some algebra to prove.
To proceed we use
(\ref{will}) and (\ref{lliE}) to write the link variables $d^{(-)}
(\vec{r},\hat{m})$ according to    \Eq{ d^{(-)} (\vec{r}, \hat{m})
= \exp{\biggl(-i\lambda \frac{a^2}g e^a E_i^a(\vec{r_c})\hat{m}_i \biggr)
}\;\exp{\biggl( iag
\frac{\sigma_a}{2} A_i^a(\vec{r_c})\hat{m}_i  \biggr)}\;,\qquad
\hat{m} > 0 \;,\label{ditAE}}
where here we find it more convenient to evaluate $A$ and $E$ at
the central point $\vec{r_c}$ of the link $ (\vec{r}, \hat{m})$.
We shall assume that
(\ref{ditAE}) is valid  for $\hat{m} > 0$.  Due to (\ref{dpid2})
it cannot then also be valid for
 $\hat{m} < 0$.  Instead, from (\ref{dpid2}) we get that
 \ba {d^{(-)}
(\vec{r}, \hat{m})}& =&d^{(-)}(\vec{r}+a\hat{m}, - \hat{m})^{-1} \cr
&=&\exp{\biggl( iag
\frac{\sigma_a}{2} A_i^a(\vec{r_c})\hat{m}_i  \biggr)}\;
 \exp{\biggl(-i\lambda \frac{a^2}g e^a E_i^a(\vec{r_c})\hat{m}_i \biggr)
}\;,\qquad
\hat{m} < 0 \;.\label{ditAEp} \ea
  We now consider the plaquette $\Gamma_0$ in the $1-2$
plane centered at $\vec{x} =(x_1,x_2)$.  Upon substituting (\ref{ditAE})
and (\ref{ditAEp})
into (\ref{dwl}) and taking $\vec{r}=(x_1 - {a\over 2},
x_2 - {a\over 2})$, $\hat{m}=(1,0)$ and   $\hat{n}=(0,1)$,
we get the following expression for $W^\lambda(\Gamma_0)$:
 \ba    W^\lambda( \Gamma_0)&=&  \Tr
 \exp{\biggl(-i\lambda \frac{a^2}g e^aE_1^a(x_1,x_2-a/2)\biggr) }
\;\exp{\biggl( iag \frac{\sigma_a}{2} A_1^a (x_1,x_2-a/2)
 \biggr)}\cr  & &\times \;
 \exp{\biggl(-i\lambda \frac{a^2}g e^aE_2^a(x_1+a/2,x_2) \biggr)
}\;\exp{\biggl( iag \frac{\sigma_a}{2} A_2^a (x_1+a/2,x_2)
 \biggr)}\cr  & &\times   \;
\;\exp{\biggl(-iag \frac{\sigma_a}{2} A_1^a (x_1,x_2+a/2)
 \biggr)}
 \exp{\biggl(i\lambda \frac{a^2}g e^a E_1^a(x_1,x_2+a/2) \biggr)}
 \cr   & &\times     \;
}\;\exp{\biggl(-iag \frac{\sigma_a}{2} A_2^a (x_1-a/2,x_2)
 \biggr)}
 \exp{\biggl(i\lambda \frac{a^2}ge^a E_2^a(x_1-a/2,x_2)\biggr)
 \;,\ea
where we used  $d^{(+)}(\vec{r}, \hat{n})^\dagger
=d^{(-)}(\vec{r}+a\hat{n}, -\hat{n})$ which follows from
(\ref{dpid1}) and (\ref{dpid2}).
After expanding this expression to fourth order in $a$,
some work shows that we recover
the Kogut-Susskind result, i.e.  \Eq{
W^\lambda(\Gamma_0)\rightarrow \Tr \biggl(
\BI  - \frac12  g^2 a^4 (F_{12}(\vec{x}))^2 \biggr)
 \;,\quad {\rm as}  \;   a\rightarrow 0\;,} which is the same as
(\ref{tsr}).  Hence $ H_{1}(\lambda)$
yields the magnetic field energy
  \Eq{     H_{1}(\lambda)\rightarrow   \frac12
\sum_{\vec{r}} \;a^3B^a_i(\vec{r}) B^a_i(\vec{r}) \;,\quad{\rm
as}\;a\rightarrow 0\;,}  and the total Hamiltonian
 $H(\lambda)$ reduces to that of $SU(2)$
Yang-Mills theory in the continuum limit.

\section{Quantization}

We now  consider the quantization of the lattice theory discussed
 in the previous section.   When Lie-Poisson symmetries are
present in the classical theory
the standard practice is to apply the method
 of deformation quantization  \cite{bffls}, where
 we do not identify the quantum mechanical
commutation relations with $i\hbar$ times the corresponding
classical Poisson brackets, but only demand that they agree
in the limit $\hbar\rightarrow 0$.  Also we do not identify   the
quantum Hamiltonian ${\bf H}(\lambda)$ function
  (with classical variables replaced by their corresponding
  quantum operators)
 with the classical Hamiltonian function $H(\lambda)$.
We instead only demand that ${\bf H}(\lambda)$
 reduces to $H(\lambda)$
in the limit $\hbar\rightarrow 0$.

Fixing  the quantum
dynamics using the method of deformation quantization requires
writing down a star product on the space of classical observables.
\cite{bffls}.  This is generally a difficult task and shall not
be attempted here.  Instead, we shall only demand that
the $SU(2)$ Poisson Lie group gauge symmetry which is present in the
classical theory gets replaced upon quantization
by a  gauge symmetry which is
associated with the quantum group $SU_q(2)$ \cite{TT}.
The latter can be defined in terms of $2\times 2$
matrices $\{T\}$ whose matrix elements $T_{ij}$ are not c-
numbers.  Rather they satisfy the commutation relations:
\Eq{ R \Tone\Ttwo=\Ttwo \Tone R  \;,\label{RTT}   }
 with $\Tone=T\otimes \BI$, $\Ttwo=\BI\otimes T$ and $R$, the quantum R-
matrix, given by
\Eq{  R=q^{-1/2}\pmatrix{q & & & \cr & 1 & & \cr & q-q^{-1} &1 & \cr
& & & q \cr}\;,     }   where $q$ is a c-number.
In addition to (\ref{RTT}), $T$ satisfies a unitarity condition
$T^{\dagger} T= \BI$ and also a deformed unimodularity condition
$det_q T=1$,  where $det_q T=T_{11}T_{22}
-qT_{12}T_{21}$.  The latter constraint
is possible because $det_q T$ so defined
commutes with all matrix elements $T_{ij}$.
$R$ satisfies the quantum Yang-Baxter equation.  If we set
$q=e^{\hbar\lambda / 2}$, then in the limit $\hbar\rightarrow 0$
 $R$ tends to $ \BI - i \hbar r + O(\hbar^2)$, and consequently
(\ref{RTT}) reduces to  \Eq{
[\;\Tone, \Ttwo\;] =i\hbar[\;r\;,\;\Tone \Ttwo\;]+ O(\hbar^2) \;.}
We thereby recover the algebra (\ref{pbva}) of the $SU(2)$ matrices
$v_A$  parametrizing the Lie-Poisson symmetries.

In  quantizing  the lattice theory described in the
previous section, we would like to define
an $SU_q(2)$ matrix $T(\vec{r})$
at each point $\vec{r}$ on the lattice.  Then the quantum analogue
of the Poisson brackets (\ref{pbvrvr}) is
\Eq{ R \Toner\Ttwor=\Ttwor \Toner R  \;,\label{RTrTr}   }
and we assume that  $SU_q(2)$ matrices at different points commute.

With regards to the $SL(2,C)$ matrix $d^{(-)}$, we replace
it by the $2\times 2$ matrix  $D^{(-)}$ having operators as
matrix elements.  The quantum analogues of the Poisson brackets
(\ref{pbdp}) and (\ref{pbdpm})  are\cite{mss}
\ba
R^{(+)} \Dmone\Dmtwo&=&\Dmtwo\Dmone R\;,\label{RDD1}\\
R^{(-)} \Dmone\Dptwo&=&\Dptwo \Dmone R\;, \label{RDD2}
\ea
 where $D^{(+)}={{D^{(-)}}^\dagger}^{-1}$,
 $R^{(+)}=R^T$ (the superscript $T$ denoting transpose) and
$R^{(-)}=R^{-1}$.
The matrices $D^{(\pm)}$ along with the commutation relations
 (\ref{RDD1}) and (\ref{RDD2}) define the {\it quantum double}.
In the limit $\hbar\rightarrow 0$,
$R^{(+)}$ and $R^{(-)}$ tend to $\BI- i\hbar r^T + O(\hbar^2)$ and
$ \BI+i\hbar r+O(\hbar^2)$, respectively, and the algebra
given in (\ref{pbdp}) and (\ref{pbdpm})
 is recovered from (\ref{RDD1}) and (\ref{RDD2}) to first order
in $\hbar$.
In addition, the commutation relations
(\ref{RDD1}) and (\ref{RDD2}) are covariant under left and
right $SU_q(2)$ transformations\cite{mss}:
\Eq{ D^{(-)} \rightarrow T_L^\dagger\; D^{(-)}\; T_R\;,\quad T_L
,T_R\in SU(2)\;.   \label{qgtr}}
Here both $T_L$ and $T_R$ satisfy commutation relations (\ref{RTT})
and we assume that matrix elements of $T_L$ commute with
 matrix elements of $T_R$.

In  the lattice theory, we  assign
a matrix $D^{(-)}(\vec{r},\hat{m})$
to each link $(\vec{r},\hat{m})$ on the lattice.
  Then the quantum analogues of the Poisson brackets
(\ref{pbldp}) and (\ref{pbldpm}) are
\ba
R^{(+)} \LDmone\LDmtwo&=&\LDmtwo\LDmone R\;,\label{RDLD1}\\
R^{(-)} \LDmone\LDptwo&=&\LDptwo \LDmone R\;, \label{RLDD2}
\ea
 where $D^{(+)}(\vec{r},\hat{m})={{D^{(-)}}(\vec{r},\hat{m})
 ^\dagger}^{-1}$,
and we assume that  $D$ matrices associated with
 different links commute.
Now gauge transformations on the quantum double variables are
given by
\Eq{ D^{(-)}(\vec{r}, \hat{m})   \rightarrow T(\vec{r})^\dagger  \;
 D^{(-)}(\vec{r}, \hat{m}) \; T(\vec{r}+a\hat{m}) \;,   \label{qlgtr}}
and as they are of the same form as
(\ref{qgtr}) they preserve  the
commutation relations [eqs. (\ref{RDLD1}) and (\ref{RLDD2})].
The commutation relations for
$D^{(\pm)}(\vec{r}, \hat{m})$ are therefore covariant under
$SU_q(2)$ gauge transformations.

We next must write down the quantum analogues of $H_0(\lambda)$
and $H_1(\lambda)$.  Our requirements are that these terms
 are invariant
under $SU_q(2)$ gauge transformations (\ref{qlgtr}) and also
that they reduce to
$H_0(\lambda)$ and $H_1(\lambda)$ when $\hbar \rightarrow 0$.

We begin with $ H_0(\lambda)$.  It represents the sum of kinetic
energies of the rotators.  Its quantum analogue
 ${\bf H}_0(\lambda)$ is known\cite{mss}.  To write it we need to
introduce the ``quantum" trace $\Tr_q$. \cite{tr}
${\Tr}_q$ of a $2\times 2$ matrix
 $M= [ M_{ij} ]$ is defined according to
\Eq{{\Tr}_q \;M=q M_{11}+ q^{-1}M_{22}\;. }
Unlike the usual trace, ${\Tr}_q$ does not have the general
property of invariance under cyclic permutations.  It does however
serve as an ``adjoint invariant" for
$SU_q(2)$.  By this we mean the following:
\Eq{{\Tr}_q \;T^{-1} M T = {\Tr}_q \;M\;, \label{TMTM}  }
where $T$ satisfies the commutation relations (\ref{RTT}) and
we assume that matrix elements of $T$ commute with those
of $M$.  The relation (\ref{TMTM})
can be explicitly verified using the $2\times2$
representations for $T$.  From (\ref{TMTM}) it follows that
$$ {\Tr}_q  \;    D^{(-)}(\vec{r}, \hat{m})
 D^{(-)}(\vec{r}, \hat{m})^\dagger  $$
is invariant under $SU_q(2)$ gauge transformations (\ref{qlgtr}).
 Then a possible choice for ${\bf H}_0(\lambda)$ is
 \Eq{ {\bf H}_{0}(\lambda)={{g^2}\over {2a\lambda^2}}
  \sum_{\vec{r},\hat{m}>0}
\biggl(  {\Tr}_q\; D^{(-)}(\vec{r}, \hat{m})
 \; D^{(-)}(\vec{r}, \hat{m})^\dagger\;-\;2\biggr)  \;,\label{qlham} }
as it reduces to (\ref{lham}) when $\hbar\rightarrow 0$.

Using the quantum trace it is also easy to write down a quantum
analogue of the Wilson loop
variables   $W^\lambda(\Gamma_{(\vec{r}, \hat{m},\hat{n})})$ defined in
(\ref{dwl}).  We write
\Eq{{\bf W}^\lambda( \Gamma_{(\vec{r}, \hat{m},\hat{n})})=
{\Tr }_q\;D^{(-)}(\vec{r}, \hat{m})\;
 D^{(-)}(\vec{r}+a\hat{m}, \hat{n}) \;
D^{(-)}(\vec{r}+a\hat{m}+ a\hat{n},-\hat{m}) \;
 D^{(-)}(\vec{r}+a\hat{n},-\hat{n})   \;. \label{Dwl}}
From (\ref{TMTM}) it  follows that
${\bf W}^\lambda(\Gamma_{(\vec{r}, \hat{m},\hat{n})})$
is invariant under $SU_q(2)$ gauge transformations (\ref{qlgtr}).
The quantum version of the Hamiltonian (\ref{dksh}) is then
\Eq{{\bf H}(\lambda)={\bf H}_0(\lambda) +{\bf H}_1(\lambda)
\;,\qquad {\bf H}_1(\lambda)
=\frac 1{ag^2}\sum_{\boxit}
\biggl( {\bf W}^\lambda( \Gamma_{(\vec{r}, \hat{m},\hat{n})})+
{\bf W}^\lambda( \Gamma_{(\vec{r}, \hat{m},\hat{n})})^\dagger -4\biggr)
 \;.\label{qdksh}}
The sum in ${\bf H}_1$ is again over all plaquettes.

\bigskip
\bigskip

{\parindent 0cm{\bf Acknowledgements:}}
We would like to thank Y. Gunal for useful
discussions.
We would also like to thank G. Marmo and P. Michor and the
members of the Schr\"odinger Institute, where this work was
started, for their hospitality.  A.S. was supported in part
by the Department of Energy, USA, under contract number
DE-FG05-84ER40141.

\end{document}